# Kinetic trapping of charge-transfer molecules at metal interfaces

Anna Werkovits,[1*] Simon Hollweger, [1*] Max Niederreiter,[2] Thomas Risse,[3] Johannes J. Cartus,[1] Martin Sterrer,[2] Sebastian Matera,[4] and Oliver T. Hofmann[1]

1 Institute of Solid State Physics, Graz University of Technology, Petersgasse 16/II, 8010 Graz, Austria

2 Institute of Physics, University of Graz, Universitätsplatz 5, 8010 Graz, Austria

3 Institut für Chemie und Biochemie, Freie Universität Berlin, Arminallee 22, 14195 Berlin, Germany

4 Theory Department, Fritz Haber Institute of the MPG, Faradayweg 4-6, 14195 Berlin-Dahlem, Germany

*Both authors contributed equally to this work

## ABSTRACT

Despite the common expectation that conjugated organic molecules on metals tend to adsorb in a flat-lying wetting layer, several recent studies have found strong indications for coverage-dependent transitions to upright-standing phases, which exhibit notably different physical properties. In this work, we argue that from an energetic perspective, thermodynamically stable upright-standing phases may be more common than hitherto thought. However, for kinetic reasons this phase may often not be observed experimentally. Indeed, using first principles kinetic Monte Carlo simulations, we find that the structure with lower molecular density is (almost) always formed first, reminiscent of Ostwald's rule of stages. The phase transitions to the thermodynamically stable upright-standing phase are likely to be kinetically hindered under conditions typically used in surface science (gas phase adsorption at low flux). This provides a possible explanation why they are commonly not observed. Investigating both the role of the growth conditions and the energetics of the interface, we find that the time for the phase transition is determined mostly by the deposition rate and, thus, mostly independent of the nature of the molecule.

**Introduction.** The extensive polymorphism exhibited by inorganic/organic interfaces can be both a blessing and a curse, as many properties, such as the interface dipole[1] or the charge-carrier mobilities[2] are strongly affected by the crystal structure at the interface.[3] A prototypical example for this is found in lying-to-standing phase transitions. These occur, e.g., when at low dosages molecules assume a flat-lying structure, but upon deposition of more material, the first layer re-orients into a more tightly packed, upright standing structure. Such structural changes are often accompanied with a sudden, large change of the molecules electron affinity,[1] and, consequently, the interface dipole.[4]

Generally, lying-to-standing phase transitions are found for "self-assembled monolayers",[5–7] i.e., organic molecules that interact weakly with the surface (e.g. mostly through van-der-Waals interactions and no or only a single docking group) and where intra-layer interactions dominate. For the other class of molecules, i.e. conjugated organic molecules that have multiple functional groups und which undergo charge-transfer reactions with the surface, it appears that most systems,[8–16,17–24] lack indications for upright standing structures – a few notable exceptions notwithstanding.[25–27]

Although it is conceivable that in some cases, the re-oriented (standing) phase never becomes thermodynamically stable, the absence of experimental evidence for these structures is not sufficient to conclude their thermodynamic instability. From a thermodynamic point of view, such lying-to-standing phase transitions should be very common also for molecules with strong molecule-substrate interactions (see Supplementary Material for a detailed discussion). A possible explanation, that has hitherto not received much attention, would be that the phase transition is kinetically prevented.

Experimentally, kinetic trapping is extremely difficult to address. This prompted us to conduct a joint study using growth experiments and first principles kinetic Monte Carlo simulations to investigate (1) under which growth conditions kinetically trapped phases are likely to occur, (2) how long a kinetically trapped phase would be expected to be stable before it transitions to the thermodynamic minimum, and (3) how the formation of a kinetically trapped phase depends on the nature of the organic adsorbate. kMC has been used successfully to describe the growth of organic molecules at inorganic/organic interfaces[28–34] often to predict the form and distribution of clusters or the size of critical nuclei on the surface,[35,36] to monitor the self-assembly process in and out of equilibrium,[37–39] or to study the morphology in thin film growth.[37,38] Here, we demonstrate timescales over which kinetic trapping usually occurs exemplarily on the deposition of tetracyanoethylene (TCNE) on Cu(111) employing the dispersion-corrected DFT derived kinetic parameters from earlier work[40] (and listed again in the Supporting Information). For this system, strong indications of a flat-lying to upright-standing phase transitions exist both experimentally and in theory.[41,42]

To keep the following discussion simple and general, we employ several approximations to the simulation of the growth process. First, we focus only on the layer in direct contact with the surface, neglecting adsorption in the second layer or beyond. This is motivated by the fact that on metals, the first monolayer ("wetting layer") usually closes before island or film growth is observed. Furthermore, for the system considered here we observe multilayer desorption at temperatures above 233 K, rendering them irrelevant for higher temperatures.[43] As second approximation, we neglect intermolecular interactions altogether, since here we want to conceptually study the situation where intermolecular interactions do not constitute a driving force towards upright-standing layers. Finally, for our simulations we stipulate that the molecule can only adsorb in two states, flat-lying or upright-standing, as shown in Figure 1. Reality is more complicated (there are different possible sites with slightly different energies),[42] but the restriction to two states simplifies the discussion here without changing the qualitative picture.

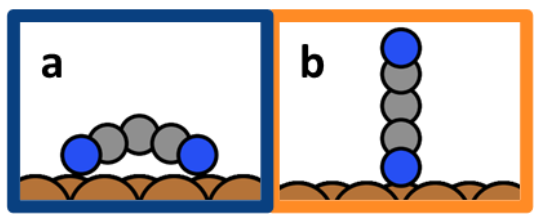

**Figure 1: a)** flat-lying **TCNE** and **b)** upright-standing **TCNE** on **Cu(111)**

Generally speaking, in thermodynamic equilibrium the most stable structure is the one that minimizes Gibb's energy per area $\gamma$:[44]

$$\gamma = \frac{1}{A}\left(\Delta F - n \cdot \mu(p, T)\right) \tag{1}$$

where $A$ is the area of the unit cell, $\Delta F$ the free energy of the structure, $\mu$ the chemical potential of the molecular reservoir depending on its pressure and temperature, and $n$ the number of molecules per unit cell. For TCNE/Cu(111), the most stable flat-lying adsorption geometry has an adsorption energy of -2.40 eV with a density of approx. 2 molecules / nm², while the most stable upright standing geometry has an adsorption energy of -1.86 eV and a footprint of approx. 4 molecules/nm².[40] The competition between the two polymorphs occurs because conjugated organic molecules can pack more densely in an upright standing adsorption geometry, but the adsorption energy is larger (more exergonic) when they maximize the contact area with the substrate, i.e. adsorb flat-lying. Consequently, the flat-lying polymorph is thermodynamically stable at low pressures and high temperatures, while the upright standing polymorph is thermodynamically preferred at high pressures and low temperatures.

Indeed, by running our Monte-Carlo simulations until equilibrium is obtained (see Method section for details), we obtain the phase diagram shown in the background of Figure 2. For the sake of discussion, we include vapor pressures that are much higher than in most experiments (up to ca. 1 bar), which would correspond to deposition rates of about 1 mm per second (see Method section for conversion). Interestingly, besides the expected phases consisting of only upright-standing molecules (shown in light orange) and of only flat-lying molecules (shown in light blue), we also find a region with both adsorption geometries (area between grey dotted lines), presumably due to configurational entropy.

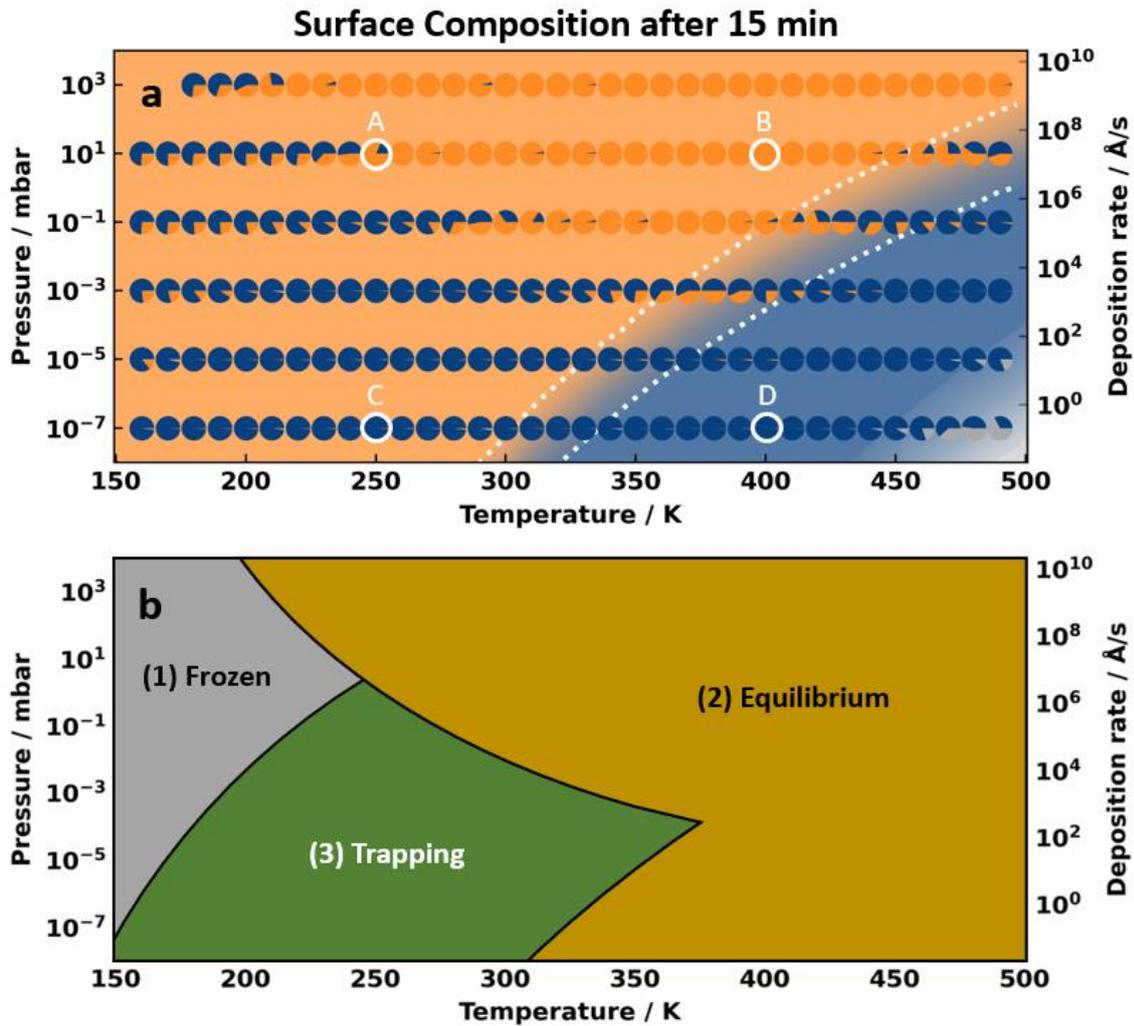

Figure 2: a) The thermodynamic phase diagram according to eq. 1 is shown in the background. In the foreground, the pie-charts represent the relative composition of the surface after 15 min have passed at the same conditions. Blue denotes the percentage of the area covered with lying molecules, orange the percentage of area covered with standing molecules, and grey empty surface. The dashed grey lines indicate qualitatively the area where a mixed composition is observed in equilibrium. For the conditions highlighted with a white circle, (A)-(D), the time evolution of the surface composition as function of time is shown in Figure 4. b) Qualitative assignment of the phase diagram into different regions (discussion see main text).

To contrast the thermodynamic expectation with the results obtained kinetically, we stop the kMC simulations after exposing the initially empty surface to the molecule reservoir for 15 minutes. The relative composition of the interface is then depicted as pie-chart in the foreground of Figure 2a. Qualitatively, we can separate the graph into three different regions: (1) At very low temperatures and higher pressures, the surface shows a mixture of standing and lying molecules, as all processes leading to changes in the orientation (including desorption) are essentially frozen. Here, the surface composition is simply determined by the way in which molecules adsorb on the surface. (Due to the symmetry of TCNE (a planar molecule with 4 peripheral CN groups), we assume that adsorbing upright-standing (on an edge) is twice as likely as flat-lying (adsorbing on its face), as discussed in more detail in the Supporting Information). This region is of no further relevance for the discussion. In region (2), which is generally found either at high pressures or high temperatures, the outcome of the kMC simulation (mostly) matches the expectation from the thermodynamic equilibrium. Finally, in region (3) we find that after a growth process of 15 min the interface consists almost entirely of flat-lying molecules, despite the thermodynamic preference of upright-standing molecules. At these conditions, the flat-lying phase is kinetically trapped. Interestingly, here, temperature alone does not seem to be the major factor, because at similar temperatures, but higher pressures / deposition rates, the stable standing phase is readily formed.

To confirm these computational expectations, we performed two sets of growth experiments. In the first experiment, we monitor the coverage of TCNE on Cu(111) using X-ray photoelectron spectroscopy (XPS), which exhibits characteristic C 1s and N 1s signals for upright-standing and flat-lying TCNE[45] (for details about the coverage determination, see Method section). We start from a clean Cu(111) surface at 300 K and expose it to different vapor pressures of TCNE until saturation at these conditions is reached. As shown in Figure 3a, at the lowest dosing pressure considered ($2 \times 10^{-10}$ mbar), we obtain a coverage of approx. 1.6 TCNE/nm², which increases quickly to 1.85 TCNE/nm² at a pressure of $10^{-8}$ mbar, and then with a smaller slope at higher dosing pressures. Notably, the coverage measured at the lowest pressure is slightly above the 1.4 TCNE/nm² reported for the closure of a full monolayer reported by Erley et al.[46] and slightly below the 2.0 TCNE/nm² we expect for a perfectly well-ordered monolayer of flat-lying molecules only.[42] While we cannot rule out completely that initially, a low-coverage phase forms that we have not considered so far (e.g., by incorporating adatoms or because of strong repulsive interactions between the molecules), our XPS analysis suggests that under these low-pressure conditions the surface is covered with disordered molecular islands consisting predominantly of flat-lying molecules with some standing molecules, and empty space in between (areal contribution of lying:standing:empty=5:3:2).[43] At higher pressures, the remaining holes in the layer are filled with upright standing TCNE molecules. This is qualitatively consistent with the computed "thermodynamic" phase diagram, which predicts mixed compositions at room temperature, moderate to low pressures, which is in agreement with the analysis of the composition at room temperature provided elsewhere.[43]

As second experiment, we exposed a freshly cleaned Cu(111) surface to TCNE at a temperature of $T =$ 200 K to form a TCNE multilayer, which was then analyzed using infra-red reflection absorption spectroscopy (IRRAS). The spectrum (blue in Fig. 3b) contains the sharp TCNE multilayer signals in the $\nu$(CN) (>2200 cm$^{-1}$) and $\nu$(CC) (<1250 cm$^{-1}$) regions, and a small, broad signal around 2170 cm$^{-1}$. In an earlier combined IR/XPS-study, we found that the latter is characteristic for flat-lying TCNE at the interface.[43] Importantly, it does not contain any indication of upright-standing molecules in significant quantities. Especially the characteristic C=C vibration of upright standing, singly-charged TCNE at 1375 cm$^{-1}$ is clearly absent. We note in passing that this does not allow to discard the presence of upright-standing TCNE, since there is an adsorption geometry with the C=C bond parallel to the surface in which, due to the surface selection rule, the intensity of this vibration theoretically vanishes.[41,42] As this signal is consistently found for full monolayer coverage at 300 K,[43] we take this as strong indication that at 200K the surface consists (almost) exclusively of flat-lying molecules. To test this hypothesis and provide a driving force out of the kinetic trapped state, we subsequently allowed the sample to reach room temperature. During the thawing of the sample, the multilayer desorbs. Re-evaluating the surface composition with IRRAS (see red line in Figure 3b), we now do find the characteristic signature of upright-standing TCNE in contact with the surface at 1368 cm$^{-1}$. This indicates that a phase transition towards the thermodynamically stable phase has partially taken place. We note that qualitatively, despite all the simplifications, our kMC simulations also suggest the presence of a partially mixed phase near these conditions (see Figure 2).

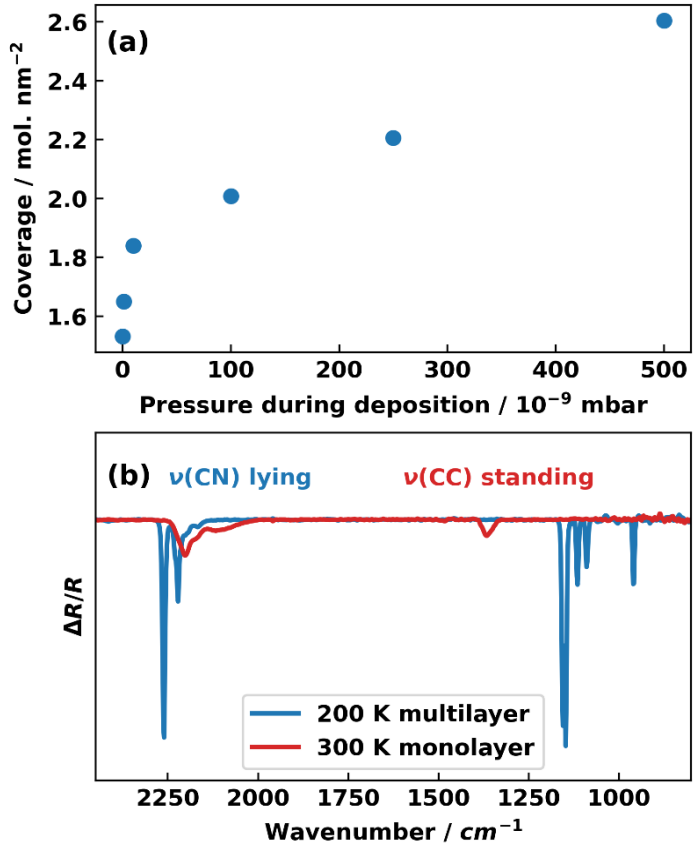

To understand why at some growth conditions thermodynamic equilibrium is reached quickly, while at others not, we analyze the simulation of the growth process and its time evolution in more detail. Initially, we start with an empty surface, on which molecules collide with the surface with a given impingement rate, given as[47]

$$k_{impingement} = s \frac{pA}{\sqrt{2\pi m k_B T}}$$

(2)

Where $p$ is the pressure, $m$ the mass of the molecule, and $A$ the area. We note in passing that flat-lying molecules occupy twice the area of a standing molecule. The sticking s is set to unity here, but adsorption is only permitted if the adjacent unit cells are still unoccupied, i.e. if there is space for the molecule to adsorb in its respective geometry. Once a molecule is adsorbed on the surface, it is free to diffuse and rotate, or to desorb. Furthermore, upright standing molecules can fall over or lying molecules can stand up. The rate constants $k$ of these processes are modelled using the Arrhenius equation

$$k = a \exp\left(-\frac{\Delta E_a}{k_B T}\right),$$

(3)

where $E_a$ is the activation energy, $a$ the attempt frequency, $k_B$ Boltzmann's constant, and $T$ the temperature. Naturally, desorption is the slowest process because the (negative) adsorption energy is much higher than the barriers for all other processes and the pre-exponential factors are similarly large (see Supporting Information). Conversely, diffusion and rotation on the surface exhibit not too large barriers and are relatively fast. The most critical rates, however, are the re-orientation processes. Because of the significantly more stable flat lying adsorption geometry, "falling over" is a much faster process than "standing up" (see Supplementary Information or ref 40).

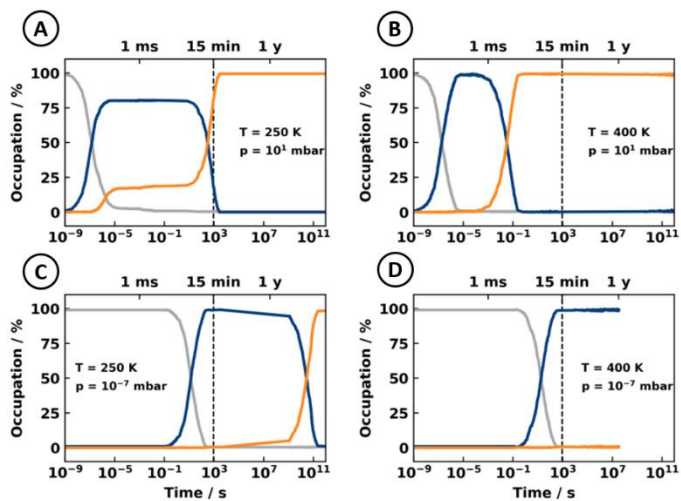

**FIGURE 4: (A)-(D)** EVOLUTION OF THE SURFACE COMPOSITION DURING THE MONTE-CARLO SIMULATION AT DIFFERENT CONDITIONS HIGHLIGHTED IN **FIGURE 2** (SEE INSET).; TRACES SHOW THE FRACTION OF SURFACE SITES BEING EMPTY (GREY TRACE), OCCUPIED BY FLAT LYING MOLECULES (BLUE), AND OCCUPIED BY MOLECULES IN AN UPRIGHT GEOMETRY (ORANGE).

The simulation shows that the growth process occurs in two stages. In the initial phase of the growth every molecule that adsorbs upright standing can diffuse on the surface and find a free spot to "fall over" to minimize its energy. The reverse process, molecules standing up, is comparatively slow and quickly undone by the same molecule falling over again. Thus, in the first stage of the growth process exclusively flat-lying molecules occupy the surface, since this is the phase minimizing the total free energy of the system.

Once the surface is completely covered with molecules, the second stage of the growth process starts. This is characterized by the joint process of a flat-lying molecule standing up and another molecule adsorbing next to it before it falls over again. For realistic (i.e., not too high) pressures and (not too low) temperatures, the rate of falling over is larger than that of adsorption, and, hence, the probability for this joint process is relatively small compared to "fluctuating", i.e., standing up and falling over without adsorption of an additional molecule. As will become important later, we find for essentially all deposition conditions that the limiting factor in this joint process is the rate of adsorption. This is the case even at very high pressures and simply a consequence of the fact that the barrier for the molecule to fall over is very small.

Although adsorption of a standing molecule in the second stage is a rare process, once it does occur it is hardly undone. None of the two standing molecules can fall over since the adjacent sites are already occupied with other molecules. The only pathway towards lying molecules would now be to desorb one of the upright standing molecules. The process of standing up and concurrent adsorption depends both on temperature (for the first part) and pressure (for the second part), while desorption depends solely on temperature (and the adsorption energy of the standing molecules).

Interestingly, the two stages of growth observed here are reminiscent of Ostwald's rule of stages,[48] which states that the phase most closely resembling the "mother phase" forms first before the thermodynamically stable phase forms. Although technically, a "mother phase" here does not exist, we find that here the phase with the lowest density, which also corresponds to the lowest energy per molecule, inevitably forms first before a thermodynamically more stable phase is formed. We note that, qualitatively, the growth behavior and the formation of a lying phase before a standing phase is the same (without intermolecular interactions) as in self-assembled monolayers (with weak molecule-substrate interactions).[39]

The different dependence of the processes on pressure and temperature allows us to rationalize the observed behavior after 15 min. Figure 5(A-D) track the surface composition as function of time for different deposition conditions. As a joint feature in all these plots, we find that initially, the flat-lying phase forms. If both temperature and deposition rate are high (shown in Figure 5b), the lying molecules often attempt to stand up. At these high background pressures, the high availability of molecules in the gas phase leads to the adsorption of upright-standing molecules. This leads to a very quick phase transition to the thermodynamically stable standing phase in a matter of seconds. Qualitatively, the situation remains similar at lower temperatures (250 K, shown in Figure 5a). Here, molecules standing up occurs less frequently, but again, once they do, another standing molecule is (irreversibly) adsorbed. Due to the lower temperature, it takes approximately 1 hour until the phase transition is completed. It is important to note, however, that the deposition rates assumed in Figure 2b and 2a are very high. At 250 K and more realistic deposition rates (ca 0.1 Å/s), the probability of the joint process of standing up and adsorbing a molecule becomes very low, such that the phase transition only occurs after several years (Figure 5c). This effectively leads to the kinetic trapping of the lying phase. Finally, when remaining at such low deposition rates but going to higher temperatures (Figure 5d), the reverse process (standing up and desorption of this molecule, leaving a sufficiently large empty space for flat-lying molecules to adsorb) starts to compete with the joint process of standing up and adsorption of another standing molecule. When the former becomes dominant, the transient population of upright-standing molecules becomes zero. This is tantamount to the flat-lying phase being thermodynamically more stable, and hence, no phase transition to an upright standing phase occurs at all. Note that at even higher temperature, also the flat-lying molecules start to desorb, leading surfaces that are only partially covered or entirely molecule-free.

An important insight from these considerations is that the limiting factor for the phase transition is mostly the pressure in the gas phase, i.e. availability of additional molecules. This implies that, the relative adsorption energies notwithstanding, it does not depend on the nature of molecule and substrate. In fact, even the height of the barriers should play only a minor role, if any. To test this statement, we repeated the kMC simulations with modified barriers for diffusion on the surface and re-orientation between standing and lying. As we show in the Supporting Information, increasing the barriers for diffusion (by up to 300 meV) has no discernable impact on the growth kinetics (see Figure S4). Maybe more surprisingly, even the barrier governing the re-orientation only has a minor impact for most realistic growth conditions. Figure 6 compares the time until the standing phase is formed for the "regular" barrier for re-orientation (580 meV to stand up, 40 meV to fall over) with a barrier that is increased by 320 meV (i.e., to 900 meV to stand up and 360 meV to fall over). For deposition rates below ca. 100 Å/s, the time until thermodynamic equilibrium is almost unaffected (within the approximations of the kMC-simulations) by the increased barriers. Only at very low temperatures and/or high deposition rates, the re-orientation becomes the rate-limiting step, i.e. becomes slower than the (re-)adsorption of molecules, as shown in Figure 6b. In this context, it is worthwhile noting that Arefi et al. showed that the re-orientation barriers generally decrease for larger molecules.[49] Since our model molecule TCNE is rather small (10 atoms), we therefore expect that adsorption is the limiting factor for the growth processes of most organic molecules.

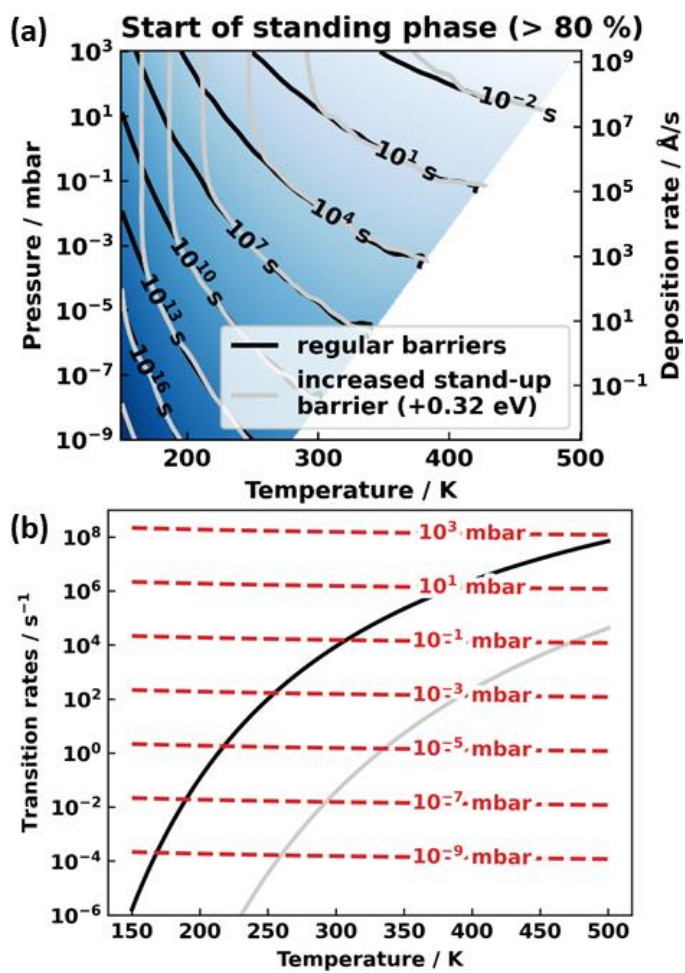

**Figure 5:** (a) Time to thermodynamic equilibrium of the standing phase, defined by a surface composition of more than 80% standing molecules. Black lines denote the barriers determined for TCNE, grey lines show the situation for a re-orientation 320meV-increased re-orientation barrier. (b) Comparison of impingement rate (red dashed line) at different pressures with the re-orientation rate (standing up) using the barriers for TCNE/Cu (black line) and barriers that are 320 meV larger (grey line).

**Summarizing,** we performed first-principles kinetic Monte Carlo simulations to study the growth of functionalized conjugated organic molecules on metal surfaces. Such molecules are likely to exhibit (at least) two different phases, a flat-lying and an upright-standing geometry. Reminiscent of Ostwald's rule of stages we find that growth generally occurs in two stages: First, a low-density phase with the higher adsorption energy per molecule is formed, before eventually the phase transition to the thermodynamically stable phase occurs. For realistic growth conditions, it appears that the limiting factor for the phase transition is the adsorption of additional molecules. Indeed, at low temperatures, we only find experimental indications for flat-lying molecules on the surface, even though a phase consisting of standing molecules should be thermodynamically preferred. Heating the sample to room temperature reduces this kinetic hindering, and under these conditions we do find a mixture of both standing and lying molecules. However, the phase transition does not complete during the time of the experiment. This is consistent with our simulations, which show that the time required to establish thermodynamic equilibrium exceeds several hours even at nominal deposition rates of 100 Å/s.

Interestingly, the obtained results are only weakly dependent on the barriers for diffusion and re-orientation. This indicates that the conditions usually employed to grow metal/organic interfaces in surface science experiments (room temperatures and growth rates below 1 Å/s, can readily lead to kinetically trapped phases and may be a reason why upright-standing layers in direct contact with metal surfaces are rarely observed.

**Methods. Computational Details.** We model the microkinetic behavior by *kinetic Monte-Carlo (kMC)* simulations via the *kmcos* code.[50] The ingredients required are the molecular adsorption sites with their kinetic interconnections. The latter are manifested through the elementary processes, our prototypical molecules undergo in an PVD experiment, i.e., adsorption/desorption, diffusion, reorientation. In detail, elementary processes are quantified spatially through their initial and final adsorption site, as well as temporally through the process rate. According to the *Variable Step Size Method*[51–53] at each step one process is randomly drawn (processes weighted by their process rates) and executed after the simulation time is forwarded by a random number distributed according to the Poisson statistic of the total rate of processes possible at the kMC step.

For this study TCNE/Cu(111) serves as prototypical system. Sorption processes are treated via collision-theory rate constants,[44] whereas on-surface processes are described via harmonic transition state theory as extensively discussed in our previous work,[40] where we applied dispersion-corrected density-functional-theory (PBE+TS$^{surf}$, 7 Cu-layers, repeated slab-approach). Based on this study, we identify 42 different adsorption geometries and 78 on-surface transitions, where symmetry equivalents are already included for both. To tackle this huge model complexity while retaining a qualitative representative description, adsorption geometries and lattice are simplified: Lying molecules have a 2x2 footprint and standing molecules 1x2 or 2x1 footprints on a square lattice with a lattice constant of 3.4 Å. This reduction to 3 adsorption geometries (Figure S2) reduces the number of on-surface transitions (Figure S3) to 24. These consist of 4 processes for lying diffusion, 12 processes for standing diffusion (translation and rotation) and 8 processes for reorientation (lying down and standing up). Along with the 3 adsorption and 3 desorption processes (one per adsorption geometry) 30 transition processes are implemented in the kMC model. Details of the approximation, as well as the used transition rate constants are stated in the Supplementary Information.

To circumvent the common time disparity problem, we applied the time acceleration method by Dybeck et al.[54,55] As discussed above, intermolecular interactions are neglected. All simulations are conducted on a lattice with 20x20 sites, starting with an empty surface. To incorporate statistic effects, the simulations at one distinct ($p,T$)-point are repeated are repeated 5 times with different random seeds. Phase diagrams are sampled by ($p,T$)-points with temperature steps of 10 K and pressure steps of 2 powers of ten mbar.

The conversion from pressure to the deposition rate $r$ in Å/s as indicated on the secondary axis of Figure 2 is

$$r(p) = k_{Adsorption}(T = 300\,K, p)\frac{m}{\rho A}$$

where $k_{Adsorption}$ is the adsorption rate from equation 1, $m$ is the mass of the molecule and $\rho$ is the mass density of the molecule in the bulk. We keep the temperature constant at 300 K since the deposition rate only varies slowly with the inverse square root of the temperature.

**Experimental Details.** The set-up used for the experiments consists of a UHV chamber system equipped with a low energy electron diffraction (LEED) apparatus, a dual-anode (Al, Mg) X-ray source and a hemispherical electron analyzer for X-ray photoelectron spectroscopy (XPS) and an attached Bruker Vertex 80v FTIR spectrometer with external mercury-cadmium-telluride (MCT) detector for infrared reflection absorption (IRRAS) spectroscopy. The Cu(111) single crystal was cleaned by repeated cycles of Ar[+] ion bombardment and annealing at 850 K until a clean and well-ordered surface was obtained, as checked with LEED and XPS. TCNE was dosed from an evacuated glass vessel through a leak valve. Prior to the dosing experiments, TCNE was resublimated to increase its purity. The XPS measurements were performed at normal emission. A previously reported XPS fitting procedure has been adapted to obtain the distribution of flat-lying and upright standing TCNE molecules in the monolayer from the C 1s and N 1s spectra.[45] For IRRAS measurements, the resolution was set to 4 cm[-1] and between 100 and 500 scans were accumulated for one spectrum. The highest TCNE monolayer coverage, as determined from the C 1s and N 1s XPS signal intensities, was obtained by first adsorbing a multilayer coverage of TCNE at 200 K, followed by warming the sample to room temperature. This coverage is referred to as 1 ML in this study.

**Acknowledgements.** Funding through the projects of the Austrian Science Fund (FWF): Y1175 and I5170-N is gratefully acknowledged. Computational results have been achieved in part using the Vienna Scientific Cluster (VSC). MN and TR are grateful to the University of Graz for financial support through the DocAcademy NanoGraz and a mobility grant, respectively.

# Supporting Information for

## Kinetic trapping of charge-transfer molecules at metal interfaces

Anna Werkovits,[1*] Simon Hollweger,[1*] Max Niederreiter,[2] Thomas Risse,[3] Johannes J. Cartus,[1] Martin Sterrer,[1] Sebastian Matera,[4] and Oliver T. Hofmann[1]

1 Institute of Solid State Physics, Graz University of Technology, Petersgasse 16/II, 8010 Graz, Austria

2 Institute of Physics, University of Graz, Universitätsplatz 5, 8010 Graz, Austria

3 Institut für Chemie und Biochemie, Freie Universität Berlin Arminallee 22

4 Theory Department of the Fritz-Haber Institute, Faradayweg 4-6, 14195 Berlin-Dahlem, Germany

*Both authors contributed equally to this work

## Thermodynamic Stability

Although most organic/metal interfaces display a flat-lying first layer, it is useful to briefly consider the relative stabilities of flat-lying viz-a-viz upright-standing geometries to estimate under which conditions we would expect the latter to be thermodynamically more favorable. Because organic molecules can be rather complex, the discussion here must be simplified. Specifically, here, we consider organic molecules that are typical for applications in organic electronics and consist of a π-conjugated core (shown in Figure 1a-c in blue) and functional groups such as cyano or carbonyl groups at its periphery (shown in Figure 1a-c in orange). We furthermore assume that the organic molecules only interact mostly via van-der-Waals forces with each other, i.e. assume that no strong, directional bonds (e.g., mediated by surface ad-atoms or through hydrogen bonds) exist.

Following (ab-initio) thermodynamics,[56] the stable structure is the one which minimizes the Gibbs energy per area $\gamma$

$$\gamma = \frac{n}{A}\left(\frac{\Delta E}{n} - \mu(T, p)\right) \qquad (2)$$

where $n/A$ is the number of molecules per area, $\Delta E/n$ the formation energy per molecule, and $\mu$ the chemical potential of the molecular reservoir, which increases with increasing pressure and decreases with increasing temperature. Generally speaking, equation 1 implies that at low chemical potentials, the structure with the best energy per molecule is thermodynamically stable, while higher chemical potentials provide a driving force towards more densely packed structures.

To discuss the energetic competition between flat-lying and upright standing layers, it is useful to conceptually separate $\Delta E$ into the contributions arising from molecule-substrate interactions and from molecule interactions, respectively. When the molecule is oriented with the $\pi$-plane parallel to the surface, i.e., lies face-on as shown in Figure 1a, all atoms of the molecule are in direct contact with the surface and interact via van-der-Waals forces (indicated as light blue arrows). As a rule of thumb, the typical magnitude of van-der-Waals forces between metals and organic molecules is about 100 meV per atom.[57] Furthermore, functional groups can also create a (partially) covalent bond.[58] Typical values for covalent bonds on surfaces range between 0.5 and 2 eV, and many molecules have multiple functional groups that are in contact at the same time. In addition, also the molecules within the layer interact with each other. Flat-lying molecules have a relatively small contact area to each other, and in the absence of strong directional interactions (adatom-mediated or hydrogen bond), typical interaction energies for molecules like TCNE,[59] acenequinones[60] or anthracene-derivates[37] are ca. 100 meV per molecule. Since this is very small compared to the molecular-substrate interaction, for the purpose of this discussion we consider the molecule-molecule interaction in flat-lying layers negligible. Summarizing the considerations, for small to mid-sized molecules, $\Delta E$ on metals typically are about 2-4 eV per molecule (including, exemplarily, for TCNE on Cu(111)[42], PTCDA on coinage metals,[57] F4TCNQ on Au(111),[61,62] pyrenetetraraone on Cu(111), and many more).

Conversely, in an upright standing molecule, indicated in Figure 1b, only the atoms of the functional group are in direct contact with the surface. Most other atoms are so far away from the surface that their van-der-Waals interaction is negligible. The main contribution to the molecule-substrate-interaction is, thus, the partially covalent bond the molecule forms with the metal. Hence, because standing molecules typically have less functional groups in contact and due to the smaller van-der-Waals interaction with the surface, isolated standing molecules typically bind less strongly (have less negative $\Delta E/n$) than isolated flat-lying molecules. Typical values are between 1-2 eV (e.g. for TCNE on Cu(111),[59] pyrenetetraone on Cu(111), or biphenylthiole on Au(111)[63,64]). Furthermore, van-der-Waals interactions between two organic molecules are much smaller than between a molecule and a metal surface, certainly not enough to overpower the $\Delta E/n$ of 2-4 eV obtained above for flat-lying structures

Because the van-der-Waals interactions between two organic molecules are smaller than between a molecule and a metal surface, the upright-standing structures will only rarely have a better energy per molecule than the flat-lying structures. Thus, in the limit of low coverage/dosage/chemical potential, usually flat-lying molecules are likely to prevail. This is also consistent with the general expectation that organic molecules in direct contact with metal substrates form a so-called "wetting layer", consisting of flat-lying molecules. However, because upright-standing molecules can pack more densely, they will eventually be lower in $\gamma$ for sufficiently large $\mu$ (see Figure 1d). Thus, if we only consider monolayers (i.e., molecules in direct contact with the surface), a flat-lying to upright-standing phase transition would be expected for almost any conjugated organic molecule (with functional groups).

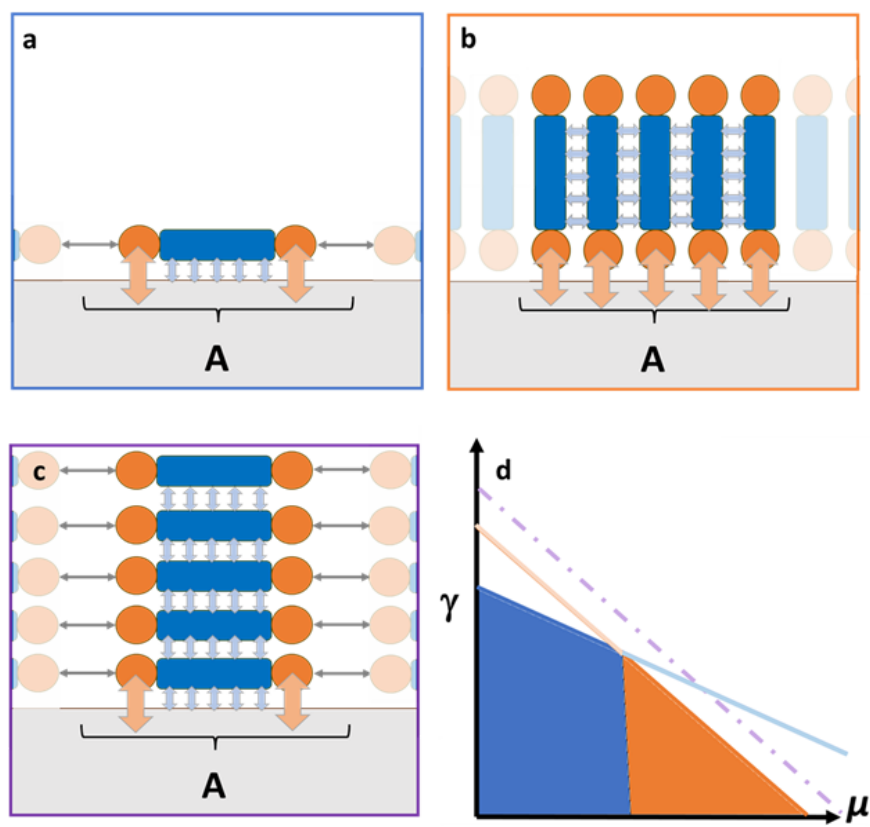

In practice, to see whether the standing phase indeed becomes stable eventually, it must be compared to a phase with a multilayer structure exhibiting the same number of molecules (schematically shown in Figure 1c). It stands to reason that the interaction between the molecules in the upright-standing monolayer and the flat-lying multilayer, which are both van-der-Waals driven, are similarly large (although, of course, differences in the geometry, the periodicity of the upright-standing monolayer parallel to the surface, and the existence of in-plane interactions in the multilayer may slightly benefit one structure over the other). Conversely, the larger difference stems from the interaction between the molecules and the surface. In a first approximation, we can expect the upright-standing phase to be more stable if the increased density of functional groups in direct contact with the surface outweighs the van-der-Waals interactions of the flat-lying molecule. Since the approximate value for the van-der-Waals interactions is 100 meV/atom (corresponding to ca. 66 meV / $Å^2$), we can expect a phase transition if additional density of functional groups at the surface for the upright standing layer is so large that this value is exceeded. Depending on the interaction energy of the functional groups (0.5 eV − 2 eV) this should be the case if the area occupied by an upright standing molecule does not exceed ca. 8-30 $Å^2$ per functional group. Although the discussion presented here is, of course, mostly qualitatively (and works with approximate magnitude of numbers, to which always exceptions can be found), we note that this condition is met for many organic molecules.

## kMC representation of adsorption geometries and transition processes

In previous studies 42 different adsorption geometries[42] and 78 on-surface transitions[40] were found in DFT-based simulations of TCNE on Cu(111). These numbers already contain all symmetry equivalents. To make such high throughput kMC simulations of this system possible, the model complexity is tremendously simplified, while the qualitative description of the kinetic effects is retained. As demonstrated in Figure S2, we approximate the adsorption geometries on the originally hexagonal Cu(111) surface by a square lattice with a lattice constant of 3.4 Å. We include only one type of the flat-lying adsorption sites (with a 2x2 footprint covering 11.56 $Å^2$) and two of the upright-standing ones (with 1x2 and 2x1 footprints, referred to as 'standing vertical' and 'standing horizontal', covering 5.78 $Å^2$). This representation also approximates the tightest packing found in ref Egger et al.[42] We note here again that intermolecular interactions are not included in this work.

As the adsorption geometries cover several lattice sites, each occupied site is treated separately in the kmcos code (different "species" in the kmcos-jargon). For example, the lying adsorption geometry is represented via the species L0, L1, L2 and L3 (Figure S2c).

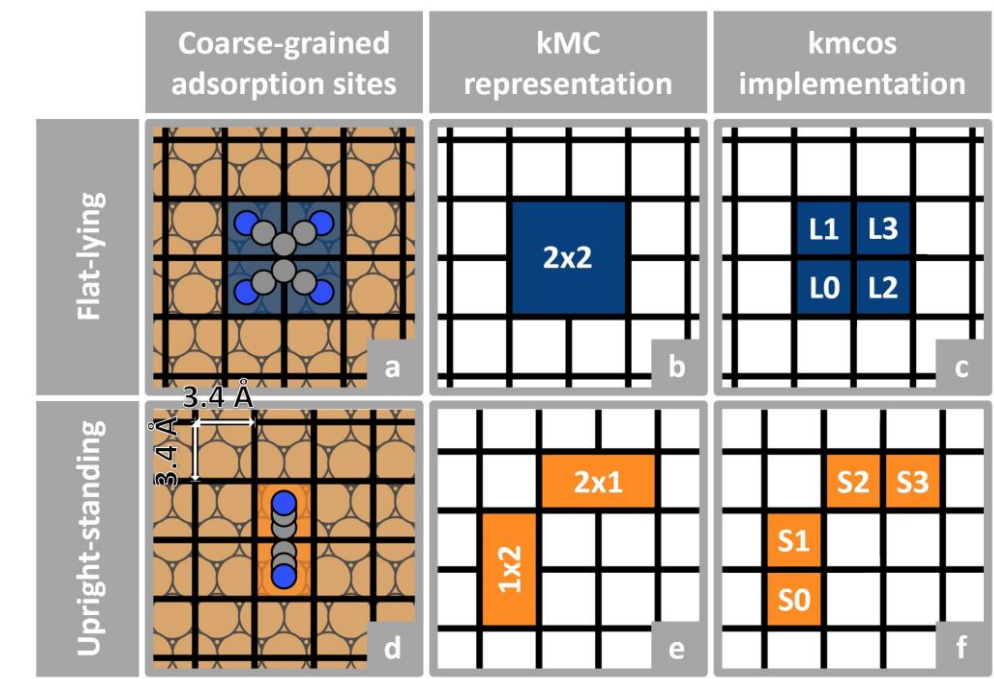

Transferring the kinetic processes found in ref[40] to the course-grained adsorption geometries, we figured out 24 on-surface processes that are implemented kMC model. In Figure S3, these are assorted in the groups lying translation, standing translation, standing rotation and standing-up and lying-down. The standing rotation only can take place if the lattice site between the horizontal and vertical orientation is unoccupied. In addition, we added for each adsorption geometry an adsorption and a desorption process – these are in total 6 sorption processes.

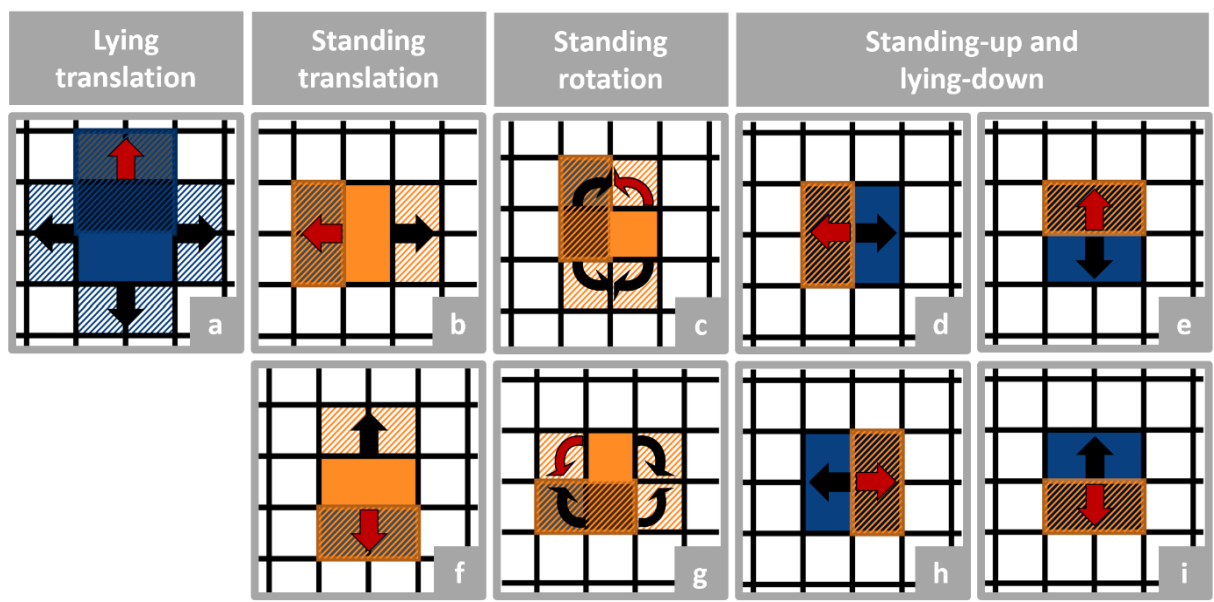

## Default barriers of kinetic processes

The rate constants of the kMC simulations are all modelled via the Arrhenius equation:

$$k = A \times \exp\left(-\frac{\Delta E}{k_B T}\right)$$

Where k is the rate constant, A the attempt frequency of the corresponding process, and ΔE the energy that must be overcome. Most of the simulations presented in the main manuscript are based on the values of TCNE/Cu(111), which we reported in an earlier paper.[40] As the number of adsorption geometries and correspondingly the number of kinetic processes is reduced, the smallest barriers (ΔE) are chosen for the streamlined process. For the sake of self-containment, we also report them here:

| Process | A [s⁻¹] | ΔE [eV] |
|---|---|---|
| **Diffusion of lying molecules** | $3 \times 10^{14}$ | 0.45 |
| **Diffusion of standing molecules** | $5 \times 10^{12}$ | 0.05 |
| **Standing up (lying → standing)** | $5 \times 10^{13}$ | 0.58 |
| **Falling over (standing → lying)** | $9 \times 10^{11}$ | 0.04 |
| **Desorption lying molecules** | Impingement factor (eq. 2) | 2.40 [barrierless] |
| **Desorption standing molecules** | Impingement factor (eq. 2) | 1.86 [barrierless] |

Note that we model the adsorption simply via the impingement factor (equation 2 in the main text), assuming a sticking coefficient of unity and a non-activated process. Since the adsorption is generally much faster than all other steps, modelling it as non-activated does not result in a loss of accuracy. The attempt frequency for the desorption is scaled by the impingement factor to account for detailed balance, as is commonly done in literature for this process.[55]

## Computational details

All the kinetic Monte Carlo simulations were performed within the kmcos simulation package.[50] The simulation was performed on a 20x20 square lattice with a side length of 3.4 Å. A time acceleration scheme (*Variable Step Size Method* [51–53]) was used to reach reasonable times at the end of the simulation. The values for the adjustable time acceleration parameters that were used in all simulation runs can be found in Table S2.

| Parameter name | Parameter value |
|---|---|
| **Buffer_parameter** | 100 |
| **Sampling_steps** | 20 |
| **Execution_steps** | 200 |
| **Threshold_parameter** | 0.3 |

For the adsorption rate we used equation 2 from the main text. This gives the total impingement rate on an area *A.* Given that we are dealing with three distinct adsorption geometries, namely 'lying', 'standing vertical', and 'standing horizontal', as previously explained, it was necessary to adjust the individual adsorption rates for these three configurations to collectively match the total adsorption rate derived from equation 2. We did this by assuming a 50/50 split in standing and lying adsorptions. To achieve this adjustment, we assumed an equal 50/50 distribution between standing and lying adsorptions. Consequently, we scaled the lying adsorption rate by a factor of 0.5, while the two standing adsorption rates were scaled by a factor of 0.25 each.

## Diffusion barrier variation

In addition to the barrier variation of the reorientation barrier, we also increased the standing diffusion barrier by 550 meV to a total diffusion barrier of 600 meV. We observed that this change does not have any remarkable impact on the growth kinetics of the standing phase as depicted in Figure S4.

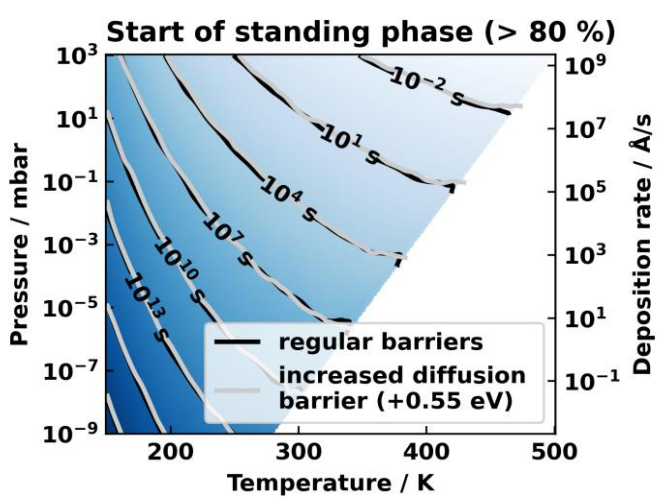

FIGURE S4: TIME TO THERMODYNAMIC EQUILIBRIUM OF THE STANDING PHASE, DEFINED BY A SURFACE COMPOSITION OF MORE THAN 80% STANDING MOLECULES. BLACK LINES DENOTE THE SAME BARRIERS AS DEFINED IN THE MAIN TEXT, GREY LINES SHOW THE SITUATION FOR A DIFFUSION BARRIER INCREASED BY 550 MEV.